# A New Energy Conservation Law for Time-Harmonic Electromagnetic Fields and Its Applications

Wen Geyi
*Research Center of Applied Electromagnetics, Nanjing University of Information Science and Technology, No.219, Ningliu Road, Nanjing, Jiangsu, China 210044*

We report a new energy conservation law for time-harmonic electromagnetic fields, which is valid for an arbitrary medium. In contrast to the well-established Poynting theorem for time-harmonic fields, the real part of the new energy conservation law gives an equation for the sum of stored electric and magnetic field energies and the imaginary part involves an equation related to the difference between the dissipated electric and magnetic energies. Universally applicable expressions for both the electric and magnetic field energies have been obtained and demonstrated to be valuable in characterizing the energy storage and transport properties in complex media. For a lossless isotropic and homogeneous medium, the new energy conservation law implies that the stored electromagnetic field energy of a radiating system enclosed by a surface is equal to the total field energy inside the surface subtracted by the energy flowing out of the surface.

Conservation laws are of fundamental importance in understanding the physical world and have found wide applications. The conservations laws specify which physical process is or is not allowed. It is believed that the physical sciences are ultimately based on various conservation laws [1]. The well-known energy conservation law for electromagnetic (EM) fields is the Poynting theorem, which has a history of more than one hundred and thirty years [2] and still plays a major role in classical electrodynamics. For time-harmonic EM fields in general medium, the Poynting theorem is given by [3]

$$-\frac{1}{2}\mathbf{E}\cdot\bar{\mathbf{J}} = \nabla\cdot\frac{1}{2}(\mathbf{E}\times\bar{\mathbf{H}}) + j2\omega\left(\frac{1}{4}\mathbf{B}\cdot\bar{\mathbf{H}} - \frac{1}{4}\mathbf{E}\cdot\bar{\mathbf{D}}\right), \quad (1)$$

where the bar denotes complex conjugate; $\omega$ is the angular frequency and $j$ is the imaginary unit. Other symbols in (1) have their conventional meanings. The Poynting theorem (1) provides an expression for the difference between the magnetic and electric field energy densities in a non-dispersive medium. However it is not clearly indicated how the electric and magnetic energy densities should be defined in general medium. Formulating the energy densities for time-harmonic fields in complex media has been an active research area for many years [4]-[16], and a common understanding is that there are no general formulations that are valid for an arbitrary medium. Many expressions of energy densities for time-harmonic fields are derived from the time-domain Poynting theorem

$$-\mathbf{J}\cdot\mathbf{E} - \nabla\cdot(\mathbf{E}\times\mathbf{H}) = \mathbf{E}\cdot\frac{\partial\mathbf{D}}{\partial t} + \mathbf{H}\cdot\frac{\partial\mathbf{B}}{\partial t}. \quad (2)$$

Since most media are temporally dispersive, it is impossible to identify the right-hand side of (2) as the derivative of an instantaneous energy density for general time-dependent fields [4][5]. It is also known that the right-hand side of (2) contains the rate of change of the stored field energy as well as the dissipation rate of the field energy, which have to be classified into two distinctive components. To derive the energy density for time-harmonic fields from (2) requires great care as different interpretations for the stored field energy and the dissipated energy may yield conflicting results. It has been a common understanding that a detailed model at the microscopic level for the medium is needed when both dispersion and dissipation are present. This means that the energy density has to be formulated separately for every medium. There have been two different approaches in this regard [6]-[14]. One is to combine the Poynting theorem (2) with the equations of motion for electric and magnetic polarizations, and the other is to use equivalent circuit models.

In this letter, we present a new energy conservation law for the time-harmonic EM fields and show that the general expressions for the electric and magnetic field energies do exist for an arbitrary medium. This overturns the aforementioned common understandings. Taking the Laplace transform of the time-domain Maxwell equations, we obtain

$$\begin{aligned}\nabla\times\mathbf{H}(\mathbf{r},s) &= \mathbf{J}(\mathbf{r},s) + s\mathbf{D}(\mathbf{r},s),\\ \nabla\times\mathbf{E}(\mathbf{r},s) &= -s\mathbf{B}(\mathbf{r},s),\end{aligned} \quad (3)$$

where $s = \alpha + j\omega$ denotes the complex frequency. For simplicity, we use $f(\mathbf{r})$ to denote the field quantity in the real frequency domain without explicitly showing its dependence on $\omega$ while the corresponding field quantity in the complex frequency domain is denoted by $f(\mathbf{r},s)$, using the same symbol in the real frequency domain but explicitly showing the dependence on $s$. From (3), we find that

$$\nabla \cdot \left[ \frac{1}{2} \mathbf{E}(\mathbf{r},s) \times \bar{\mathbf{H}}(\mathbf{r},s) \right] = -\frac{1}{2} \mathbf{E}(\mathbf{r},s) \cdot \bar{\mathbf{J}}(\mathbf{r},s)$$
$$-\frac{1}{2}\alpha \left[ \bar{\mathbf{H}}(\mathbf{r},s) \cdot \mathbf{B}(\mathbf{r},s) + \mathbf{E}(\mathbf{r},s) \cdot \bar{\mathbf{D}}(\mathbf{r},s) \right] \quad (4)$$
$$-\frac{1}{2} j\omega \left[ \bar{\mathbf{H}}(\mathbf{r},s) \cdot \mathbf{B}(\mathbf{r},s) - \mathbf{E}(\mathbf{r},s) \cdot \bar{\mathbf{D}}(\mathbf{r},s) \right].$$

For an arbitrary analytic function $f(\mathbf{r},s)$, the Cauchy-Riemann conditions imply

$$\frac{\partial f(\mathbf{r},s)}{\partial \alpha} = -j \frac{\partial f(\mathbf{r},s)}{\partial \omega}. \quad (5)$$

If $\alpha$ is sufficiently small, we may have the following first order series expansion

$$f(\mathbf{r},s) \approx f(\mathbf{r}) - j\alpha \frac{\partial f(\mathbf{r})}{\partial \omega}, \quad (6)$$

By use of (5) and (6), (4) may be rewritten as

$$\nabla \cdot \frac{1}{2} \left[ \mathbf{E}(\mathbf{r}) \times \bar{\mathbf{H}}(\mathbf{r}) \right]$$
$$+ j\alpha \nabla \cdot \frac{1}{2} \left[ \mathbf{E}(\mathbf{r}) \times \frac{\partial \bar{\mathbf{H}}(\mathbf{r})}{\partial \omega} - \frac{\partial \mathbf{E}(\mathbf{r})}{\partial \omega} \times \bar{\mathbf{H}}(\mathbf{r}) \right]$$
$$= -\frac{1}{2} \mathbf{E}(\mathbf{r}) \cdot \bar{\mathbf{J}}(\mathbf{r}) - j\alpha \frac{1}{2} \left[ \mathbf{E}(\mathbf{r}) \cdot \frac{\partial \bar{\mathbf{J}}(\mathbf{r})}{\partial \omega} - \frac{\partial \mathbf{E}(\mathbf{r})}{\partial \omega} \cdot \bar{\mathbf{J}}(\mathbf{r}) \right]$$
$$-\frac{1}{2} j\omega \left[ \mathbf{B}(\mathbf{r}) \cdot \bar{\mathbf{H}}(\mathbf{r}) - \mathbf{E}(\mathbf{r}) \cdot \bar{\mathbf{D}}(\mathbf{r}) \right] \quad (7)$$
$$-\frac{1}{2} \alpha \left[ \mathbf{B}(\mathbf{r}) \cdot \bar{\mathbf{H}}(\mathbf{r}) + \mathbf{E}(\mathbf{r}) \cdot \bar{\mathbf{D}}(\mathbf{r}) \right]$$
$$-\frac{1}{2} \alpha\omega \left[ \bar{\mathbf{H}}(\mathbf{r}) \cdot \frac{\partial \mathbf{B}(\mathbf{r})}{\partial \omega} - \mathbf{B}(\mathbf{r}) \cdot \frac{\partial \bar{\mathbf{H}}(\mathbf{r})}{\partial \omega} \right]$$
$$-\frac{1}{2} \alpha\omega \left[ \mathbf{E}(\mathbf{r}) \cdot \frac{\partial \bar{\mathbf{D}}(\mathbf{r})}{\partial \omega} - \bar{\mathbf{D}}(\mathbf{r}) \cdot \frac{\partial \mathbf{E}(\mathbf{r})}{\partial \omega} \right].$$

Comparing the coefficients of similar terms, we obtain the Poynting theorem (1) and the following new energy conservation law for an arbitrary medium

$$\frac{1}{4} \mathbf{E} \cdot \bar{\mathbf{D}} + \frac{\omega}{4} \left( \mathbf{E} \cdot \frac{\partial \bar{\mathbf{D}}}{\partial \omega} - \bar{\mathbf{D}} \cdot \frac{\partial \mathbf{E}}{\partial \omega} \right)$$
$$+ \frac{1}{4} \mathbf{B} \cdot \bar{\mathbf{H}} + \frac{\omega}{4} \left( \bar{\mathbf{H}} \cdot \frac{\partial \mathbf{B}}{\partial \omega} - \mathbf{B} \cdot \frac{\partial \bar{\mathbf{H}}}{\partial \omega} \right) \quad (8)$$
$$= -j\frac{1}{4} \nabla \cdot \left[ \mathbf{E} \times \frac{\partial \bar{\mathbf{H}}}{\partial \omega} - \frac{\partial \mathbf{E}}{\partial \omega} \times \bar{\mathbf{H}} \right] - j\frac{1}{4} \left[ \mathbf{E} \cdot \frac{\partial \bar{\mathbf{J}}}{\partial \omega} - \frac{\partial \mathbf{E}}{\partial \omega} \cdot \bar{\mathbf{J}} \right].$$

The above equation does not explicitly involve the medium parameters and can be verified by using the time-harmonic Maxwell equations. The physical significance of the new energy conservation law (8) becomes clear if it is decomposed into two equations

$$w_{se} + w_{sm} =$$
$$\nabla \cdot \mathrm{Im} \frac{1}{4} \left[ \mathbf{E} \times \frac{\partial \bar{\mathbf{H}}}{\partial \omega} - \frac{\partial \mathbf{E}}{\partial \omega} \times \bar{\mathbf{H}} \right] + \mathrm{Im} \frac{1}{4} \left[ \mathbf{E} \cdot \frac{\partial \bar{\mathbf{J}}}{\partial \omega} - \frac{\partial \mathbf{E}}{\partial \omega} \cdot \bar{\mathbf{J}} \right], \quad (9)$$

$$w_{de} - w_{dm} =$$
$$-\mathrm{Im}\, \pi\omega \left( \mathbf{E} \cdot \frac{\partial \bar{\mathbf{D}}}{\partial \omega} - \bar{\mathbf{D}} \cdot \frac{\partial \mathbf{E}}{\partial \omega} \right) + \mathrm{Im}\, \pi\omega \left( \mathbf{H} \cdot \frac{\partial \bar{\mathbf{B}}}{\partial \omega} - \bar{\mathbf{B}} \cdot \frac{\partial \mathbf{H}}{\partial \omega} \right) \quad (10)$$
$$-\nabla \cdot \mathrm{Re}\, \pi \left[ \mathbf{E} \times \frac{\partial \bar{\mathbf{H}}}{\partial \omega} - \frac{\partial \mathbf{E}}{\partial \omega} \times \bar{\mathbf{H}} \right] - \mathrm{Re}\, \pi \left[ \mathbf{E} \cdot \frac{\partial \bar{\mathbf{J}}}{\partial \omega} - \frac{\partial \mathbf{E}}{\partial \omega} \cdot \bar{\mathbf{J}} \right],$$

where

$$w_{se} = \mathrm{Re}\, \frac{1}{4} \left\{ \mathbf{E} \cdot \bar{\mathbf{D}} + \omega \left( \mathbf{E} \cdot \frac{\partial \bar{\mathbf{D}}}{\partial \omega} - \bar{\mathbf{D}} \cdot \frac{\partial \mathbf{E}}{\partial \omega} \right) \right\},$$
$$w_{sm} = \mathrm{Re}\, \frac{1}{4} \left\{ \mathbf{B} \cdot \bar{\mathbf{H}} + \omega \left( \bar{\mathbf{H}} \cdot \frac{\partial \mathbf{B}}{\partial \omega} - \mathbf{B} \cdot \frac{\partial \bar{\mathbf{H}}}{\partial \omega} \right) \right\}, \quad (11)$$

$$w_{de} = \mathrm{Im}\, \pi \mathbf{E} \cdot \bar{\mathbf{D}}, \quad w_{dm} = \mathrm{Im}\, \pi \mathbf{H} \cdot \bar{\mathbf{B}}. \quad (12)$$

We now show that $w_{se}$ and $w_{sm}$ can be interpreted as stored electric and magnetic field energy densities, while $w_{de}$ and $w_{dm}$ can be interpreted as dissipated electric and magnetic field energy densities respectively. Let us first consider an isotropic dispersive and absorptive medium with the permittivity $\varepsilon = \varepsilon' - j\varepsilon''$ and permeability $\mu = \mu' - j\mu''$. It follows from (11) and (12) that

$$w_{se} = \frac{1}{4} \varepsilon' |\mathbf{E}|^2 + \frac{\omega}{4} \left[ \frac{\partial \varepsilon'}{\partial \omega} |\mathbf{E}|^2 + 2\varepsilon'' \mathrm{Im} \left( \bar{\mathbf{E}} \cdot \frac{\partial \mathbf{E}}{\partial \omega} \right) \right],$$
$$w_{sm} = \frac{1}{4} \mu' |\mathbf{H}|^2 + \frac{\omega}{4} \left[ \frac{\partial \mu'}{\partial \omega} |\mathbf{H}|^2 + 2\mu'' \mathrm{Im}\, \bar{\mathbf{H}} \cdot \frac{\partial \mathbf{H}}{\partial \omega} \right], \quad (13)$$

$$w_{de} = \pi \varepsilon'' |\mathbf{E}|^2, \quad w_{dm} = \pi \mu'' |\mathbf{H}|^2. \quad (14)$$

It is interesting to note that the stored electric and magnetic field energy densities contain terms $2\varepsilon'' \mathrm{Im}(\bar{\mathbf{E}} \cdot \partial \mathbf{E}/\partial \omega)$ and $2\mu'' \mathrm{Im}(\bar{\mathbf{H}} \cdot \partial \mathbf{H}/\partial \omega)$ respectively, which did not appear in previous reports. The reason is that the stored field energy densities have usually been derived by using an assumption that the fields are wave-packets and their amplitudes vary very slowly with time and frequency [5]. As a result, the frequency derivatives of the fields have been neglected.

Materials of negative permittivity can be characterized by the Lorentz formula[8]

$$\varepsilon = \varepsilon_0 \left( 1 + \frac{\omega_p^2}{\omega_r^2 - \omega^2 + j\Gamma_e \omega} \right),$$

where $\omega_r$ is the resonant angular frequency of atoms or structural elements of material, $\omega_p$ is the characteristic frequency, and $\Gamma_e$ is the damping coefficient. According to (13), the stored electric field energy density is given by

$$w_{se} = \frac{1}{4}\varepsilon_0\left[1+\frac{\omega_p^2(\omega_r^2+\omega^2)}{(\omega_r^2-\omega^2)^2+\Gamma_e^2\omega^2}\right]|\mathbf{E}|^2$$
$$-\frac{1}{2}\frac{\varepsilon_0\omega^2\omega_p^2\Gamma_e^2(\omega^2+\omega_r^2)}{\left[(\omega_r^2-\omega^2)^2+\Gamma_e^2\omega^2\right]^2}|\mathbf{E}|^2 \quad (15)$$
$$+\frac{1}{2}\frac{\varepsilon_0\omega^2\omega_p^2\Gamma_e}{(\omega_r^2-\omega^2)^2+\Gamma_e^2\omega^2}\operatorname{Im}\left(\bar{\mathbf{E}}\cdot\frac{\partial\mathbf{E}}{\partial\omega}\right).$$

The last two terms on the right-hand side of (15) were missing in previous studies although they are negligible for small $\Gamma_e$. Similarly we have

$$w_{de} = \frac{\pi\varepsilon_0\omega\omega_p^2\Gamma_e}{(\omega_r^2-\omega^2)^2+\Gamma_e^2\omega^2}|\mathbf{E}|^2. \quad (16)$$

For an anisotropic medium defined by $\mathbf{D}=\ddot{\boldsymbol{\varepsilon}}\cdot\mathbf{E}$ and $\mathbf{B}=\ddot{\boldsymbol{\mu}}\cdot\mathbf{H}$ in which $\ddot{\boldsymbol{\varepsilon}}$ and $\ddot{\boldsymbol{\mu}}$ are dyads, the real part of the total energy density can be written as

$$w_{se}+w_{sm} = \frac{1}{4}\operatorname{Re}\mathbf{E}\cdot\ddot{\boldsymbol{\varepsilon}}\cdot\bar{\mathbf{E}}+\frac{1}{4}\operatorname{Re}\bar{\mathbf{H}}\cdot\ddot{\boldsymbol{\mu}}\cdot\mathbf{H}$$
$$+\frac{\omega}{4}\operatorname{Re}\left(\mathbf{E}\cdot\frac{\partial\ddot{\boldsymbol{\varepsilon}}}{\partial\omega}\cdot\bar{\mathbf{E}}+\overline{\mathbf{E}\cdot\ddot{\boldsymbol{\varepsilon}}\cdot\frac{\partial\mathbf{E}}{\partial\omega}}-\bar{\mathbf{E}}\cdot\ddot{\boldsymbol{\varepsilon}}^\dagger\cdot\frac{\partial\mathbf{E}}{\partial\omega}\right) \quad (17)$$
$$+\frac{\omega}{4}\operatorname{Re}\left(\bar{\mathbf{H}}\cdot\frac{\partial\ddot{\boldsymbol{\mu}}}{\partial\omega}\cdot\mathbf{H}+\overline{\bar{\mathbf{H}}\cdot\ddot{\boldsymbol{\mu}}\cdot\frac{\partial\mathbf{H}}{\partial\omega}}-\bar{\mathbf{H}}\cdot\ddot{\boldsymbol{\mu}}^\dagger\cdot\frac{\partial\mathbf{H}}{\partial\omega}\right),$$

where the superscript '$\dagger$' denotes the Hermitian transpose. For the lossless case, we have $\ddot{\boldsymbol{\varepsilon}}=\ddot{\boldsymbol{\varepsilon}}^\dagger$ and $\ddot{\boldsymbol{\mu}}=\ddot{\boldsymbol{\mu}}^\dagger$. Therefore the stored energy density (17) can be written as

$$w_{se}+w_{sm} = \frac{1}{4}\bar{\mathbf{E}}\cdot\frac{\partial(\omega\ddot{\boldsymbol{\varepsilon}})}{\partial\omega}\cdot\mathbf{E}+\frac{1}{4}\bar{\mathbf{H}}\cdot\frac{\partial(\omega\ddot{\boldsymbol{\mu}})}{\partial\omega}\cdot\mathbf{H}. \quad (18)$$

Equation (18) is a well-known expression and has been derived in [4][5] from (2) under the assumption that the fields are wave-packets and their amplitudes vary very slowly with time and frequency. In our derivation, no additional assumption is needed.

For a general lossless medium, we have $\operatorname{Re}(\mathbf{E}\times\bar{\mathbf{H}})=0$. This implies

$$\operatorname{Im}(\mathbf{E}\cdot\bar{\mathbf{D}}-\bar{\mathbf{H}}\cdot\mathbf{B}) = -\operatorname{Im}(\bar{\mathbf{E}}\cdot\mathbf{D}+\bar{\mathbf{H}}\cdot\mathbf{B}) = 0. \quad (19)$$

in a source-free region. Taking the frequency derivative yields

$$\operatorname{Im}\left(\bar{\mathbf{E}}\cdot\frac{\partial\mathbf{D}}{\partial\omega}+\mathbf{D}\cdot\frac{\partial\bar{\mathbf{E}}}{\partial\omega}+\bar{\mathbf{H}}\cdot\frac{\partial\mathbf{B}}{\partial\omega}+\mathbf{B}\cdot\frac{\partial\bar{\mathbf{H}}}{\partial\omega}\right)$$
$$= \operatorname{Im}\left(\bar{\mathbf{E}}\cdot\frac{\partial\mathbf{D}}{\partial\omega}-\bar{\mathbf{D}}\cdot\frac{\partial\mathbf{E}}{\partial\omega}+\bar{\mathbf{H}}\cdot\frac{\partial\mathbf{B}}{\partial\omega}-\bar{\mathbf{B}}\cdot\frac{\partial\mathbf{H}}{\partial\omega}\right) = 0. \quad (20)$$

By use of (19) and (20), we may arrive at

$$\operatorname{Re}w = \frac{1}{4}\bar{\mathbf{E}}(\mathbf{r})\cdot\mathbf{D}(\mathbf{r})+\frac{\omega}{4}\left(\bar{\mathbf{E}}\cdot\frac{\partial\mathbf{D}}{\partial\omega}-\bar{\mathbf{D}}\cdot\frac{\partial\mathbf{E}}{\partial\omega}\right)$$
$$+\frac{1}{4}\mathbf{B}(\mathbf{r})\cdot\bar{\mathbf{H}}(\mathbf{r})+\frac{\omega}{4}\left(\bar{\mathbf{H}}\cdot\frac{\partial\mathbf{B}}{\partial\omega}-\bar{\mathbf{B}}\cdot\frac{\partial\mathbf{H}}{\partial\omega}\right). \quad (21)$$

This is the most general expression for the stored EM field energy density in a lossless medium. It was obtained in [16] by means of (19), (20) and the energy density expression in the time domain but has largely been ignored.

The stored energy of an EM radiator is usually defined as the difference between the total EM field energy and the radiated energy, and has been investigated by many researchers (see [17]-[19] and references therein). Since both the total field energy and the radiated energy are infinite in an unbounded space, their difference is expected to be a finite number. The above definition is actually based on an unproven hypothesis that the infinity in the total field energy is created by the energy flow associated with radiated power, which has raised discussions[17]. Indeed this definition has never been rigorously proven to be adequate. The new energy conservation law is just right to fulfill the task. Let us consider a radiator described by the source distribution $\mathbf{J}$ confined in a finite region $V_0$ surrounded by a lossless isotropic homogeneous medium with permeability $\mu$ and permittivity $\varepsilon$. In this case, equation (8) reduces to

$$\tilde{W} = W_m^{dom}+W_e^{dom}-W_{rad}, \quad (22)$$

where

$$\tilde{W} = \frac{1}{4}\operatorname{Im}\int_{V_0}\left[\mathbf{E}\cdot\frac{\partial\bar{\mathbf{J}}}{\partial\omega}-\frac{\partial\mathbf{E}}{\partial\omega}\cdot\bar{\mathbf{J}}\right]dV, \quad (23)$$

$$W_e^{dom} = \int_V w_e^{dom}dV, \quad W_m^{dom} = \int_V w_m^{dom}dV. \quad (24)$$

$$W_{rad} = \frac{1}{4}\operatorname{Im}\int_S\left[\mathbf{E}\times\frac{\partial\bar{\mathbf{H}}}{\partial\omega}-\frac{\partial\mathbf{E}}{\partial\omega}\times\bar{\mathbf{H}}\right]\cdot\mathbf{u}_n dS. \quad (25)$$

Note that both the total energy $W_m^{dom}+W_e^{dom}$ and the surface integral $W_{rad}$ approach to infinity as the surface $S$ approaches to infinity, while their difference $W_m^{dom}+W_e^{dom}-W_{rad}$ must be a finite number given by (23), which do not change as the surface $S$ expands. Assume that the surface $S$ is a sufficiently large sphere of radius $r$. By using the far-field approximations and the Funk-Hecke formula [20], we obtain

$$W_{rad} = \frac{r}{c}P_{rad}-W_{fd}, \quad (26)$$

where $c=1/\sqrt{\mu\varepsilon}$ is the wave speed, and

$$W_{fd} =$$
$$-\frac{k^2c^2\eta}{8\pi}\int_{V_0}\int_{V_0}\operatorname{Im}\left[\bar{\rho}(\mathbf{r}')\frac{\partial\rho(\mathbf{r}'')}{\partial\omega}\right]\frac{\sin kR}{kR}dV(\mathbf{r}')dV(\mathbf{r}'') \quad (27)$$
$$+\frac{k^2\eta}{8\pi}\int_{V_0}\int_{V_0}\operatorname{Im}\left[\bar{\mathbf{J}}(\mathbf{r}')\cdot\frac{\partial\mathbf{J}(\mathbf{r}'')}{\partial\omega}\right]\frac{\sin kR}{kR}dV(\mathbf{r}')dV(\mathbf{r}''),$$

In the above, $k = \omega/c$, $\eta = \sqrt{\mu/\varepsilon}$, $R = |\mathbf{r}' - \mathbf{r}''|$ and $\rho = j\nabla \cdot \mathbf{J}/\omega$ is the charge density. The energy term $W_{fd}$ is generated by the frequency derivative of the source distributions and appears in the expression of the total stored energy previously obtained [19]. According to the energy balance relation (22), we may interpret $W_{rad}$ as the total EM field energy flowing out of the surface $S$ and $\tilde{W}$ as the total stored energy for the radiator. It is noted that (22) has also been derived in [18], where the frequency derivative $\partial \bar{\mathbf{J}}/\partial \omega$ has been set to zero by the author without going one step further to investigate the physical implication of the equation. By using the integral representation of the fields, it can be shown that the total stored energy defined by (23) agrees with the previous studies [19].

In summary, a new energy conservation law for time-harmonic fields in an arbitrary medium has been discovered, and the general expressions for both the electric and magnetic field energies have been obtained, interpreted and validated by a number of applications. These general expressions for the field energies facilitate the formulation of the energy densities in various complex media and help avoid the conflicting or erroneous results from difficulties in identifying the stored field energy and the dissipated energy in conventional approaches. The new energy conservation law lays a solid foundation for the definition of the stored energy of a radiating system and proves that the infinity of the total energy in an unbounded space is caused by the radiated energy. We believe that the new energy conservation law will have many potential applications in electrical and optical engineering, especially in the study of energy storage and transport properties in new materials. In this respect, much remains to be explored.